\documentclass[amsmath,amssymb,prb,superscriptaddress,twocolumn,showpacs,longbibliography]{revtex4-1}
\usepackage{graphicx}
\usepackage{dcolumn}
\usepackage[colorlinks=true,linkcolor=blue,citecolor=blue,urlcolor=blue]{hyperref}
\usepackage{stmaryrd}
\usepackage{esvect}
\usepackage{multirow}
\usepackage{upgreek}
\usepackage{soul}
\allowdisplaybreaks
\usepackage{gensymb}

\usepackage{bm}
\usepackage{epsfig}
\usepackage{soul}
\usepackage{subfigure}
\usepackage{multirow}
\usepackage{hhline}
\usepackage{siunitx}

\usepackage{amssymb}
\usepackage{amsfonts}
\usepackage{color}
\usepackage{dsfont}
\usepackage[normalem]{ulem}
\usepackage{dcolumn}
\usepackage{mathtools}
\usepackage{siunitx}
\usepackage[usenames,dvipsnames]{xcolor}
\usepackage{hyperref}
\usepackage{epsfig}
\usepackage{graphicx}
\usepackage{subfigure}
\usepackage{blindtext}
\usepackage{outlines}
\usepackage{enumitem}
\usepackage{soul}
\setlist[enumerate,2]{label=\roman*)}
\setlist[enumerate,3]{label=\alph*)}
\usepackage{multirow}

\usepackage{ragged2e}
\DeclareSIUnit{\rydberg}{Ry}
\DeclareSIUnit{\atomicunits}{a.u.}

\makeatother
\begin{document}
\title{Multiple magnetic states of CoPc molecule on a two-dimensional layer of  \texorpdfstring{NbSe$_2$}{TEXT}}
\author{Ana M. Montero}\email{a.montero@fz-juelich.de}
\affiliation{Peter Gr\"{u}nberg Institut and Institute for Advanced Simulation,
	Forschungszetrum J\"{u}lich and JARA, 52425 J\"{u}lich, Germany}\author{Filipe S. M. Guimar\~aes}
\affiliation{J\"ulich Supercomputing Centre, Forschungszentrum J\"{u}lich and JARA, 52425 J\"{u}lich, Germany}	
\author{Samir Lounis}\email{s.lounis@fz-juelich.de}
\affiliation{Peter Gr\"{u}nberg Institut and Institute for Advanced Simulation,
	Forschungszetrum J\"{u}lich and JARA, 52425 J\"{u}lich, Germany}	
\affiliation{Faculty of Physics, University of Duisburg-Essen, 47053 Duisburg, Germany}

\begin{abstract}
Molecular spintronics hinges on the detailed understanding of electronic and magnetic properties of molecules interfaced with various materials. Here we demonstrate with ab-initio simulations that the prototypical Co-phthalocyanine (CoPc) molecule can surprisingly develop multi-spin states once deposited on the two-dimensional 2H-NbSe$_2$ layer. Conventional calculations based on density functional theory (DFT) show the existence of low, regular and high spin states, which reduce to regular and high spins states once correlations are incorporated with a DFT+$U$ approach. Depending on $U$, the ground state is either the low spin or high spin state with energy differences affected by the molecular orientation on top of the substrate. Our results are compared to recent scanning probe measurements and motivate further theoretical and experimental studies on the unveiled rich multi-magnetic behavior of CoPc molecule.
\end{abstract}

\maketitle

\section{Introduction}

Molecular spintronics developed into a field of research with the aim of combining various intrinsic properties of molecules with spintronic concepts to unveil electronic, magnetic and transport features that are of interest for fundamental research and for possible applications in information technology~\cite{Rocha2005,Molecule17}. In this framework, transition metal phthalocyanines (TMPcs) and their derivatives were established as prototypical molecules for extensive investigations owing to their chemical and thermal stabilities, functional flexibility and structural simplicity. These molecules can be magnetic or not, and can develop zero-bias anomalies such as Kondo features, Yu-Shiba-Rusinov states, spin- and vibrational-excitations (see e.g. Refs.~\citenum{Zhao05,Gao07,Wende07,Tsukahara09,Mugarza10,Brede10,Mugarza11,Franke11,Kroll12,Mugarza12,Hatter15,Brumboiu16,Kluegel14,Kezilebieke18,Wruss19,Kezilebieke19}). 

CoPc molecule is one example of TMPc molecules, which is known to carry a spin of $1/2$ with a Land\'e g-factor of 2 in gas phase~\cite{Assour65}. This magnetic state seems to be protected once the molecule is deposited on various substrates. The divalent Co ion is however a 3d$^7$ system, which can also carry a spin of $3/2$ that defines a high spin state in contrast to the low spin $1/2$ one~\cite{Kroll09}. An enormous amount of theoretical studies were performed on the basis of various computational techniques, which usually recover the low spin state of the molecule~\cite{Figgis89,Rosa92} --- a result challenged by quantum chemistry based simulations~\cite{Kroll09}. Bhattacharjee et al.~\cite{Bhattacharjee10} investigated via density functional theory (DFT) the impact of electronic correlations within the framework of DFT+$U$~\cite{Anisimov97}. They found that two energy minima corresponding to the low and high spin states for $U$ values larger than \SI{2}{\electronvolt}. Although the ground state is mostly the low spin configuration, the energy difference with the high-spin state decreases significantly when $U$ is increased. Interestingly, the electronic structure (e.g., the orbital occupation) can dramatically change with $U$. 
Once the CoPc molecules are deposited on surfaces, the majority of theoretical studies are based on various flavors of DFT and DFT+$U$ calculations (see for example, Refs.~\citenum{Brede10,Mugarza12,Salomon13,Klar14,Liang17,Wang19}). We note, however, that DFT+$U$ is often used without necessarily discussing the impact of correlations on the electronic structure of the free molecule within the very same setup. Intuitively, one expects the electronic correlations to be more important in the free standing case, where the electrons are generally more localized than when hybridization with surface electronic states is enabled. 

Experimentally, CoPc molecule is  often found to carry a spin $1/2$, independently from the nature of the substrate. Such a spin is then subject to Kondo screening and can trigger Yu-Shiba-Rusinov (YSR) in-gap states on superconducting materials. This was recently demonstrated with scanning tunneling spectroscopy/microscopy (STS/STM) on 2H-NbSe$_2$ by Kezilebieke et al.~\cite{Kezilebieke18}. In contrast to that study, the recent STS/STM investigation of Wang et al. surprisingly revealed two different states depending on the rotation of the molecule on the same substrate~\cite{Wang20}. One configuration leads to the conventional Kondo feature accompanied by a YSR in-gap state in the superconducting phase, which is then interpreted as being in the $1/2$ spin state with a Land\'e g-factor of about $1.5$~\cite{Wang20}. The other one shows no in-gap features, advocating for a non-magnetic state. However, this state presents inelastic electronic features, which are interpreted as resulting from spin-excitations that respond to a magnetic field and the presence of Dzyaloshinskii-Moriya interactions between the moments of the ligands and Co atom. This is then understood as a transition from a spin 0 (single) to a spin 1 (triplet) state. With a g-factor of 1.5, the $1/2$ spin state would carry a magnetic moment of 0.75 $\mu_B$ while the magnetic moment corresponding to the triplet state would be 1.5 $\mu_B$ assuming  the same g-factor.

These findings motivate the present ab-initio-based study, where we prospect the magnetic states of CoPc deposited on 2H-NbSe$_2$. First, we revisit the free standing molecule using DFT+$U$. Although the molecule is found to carry a magnetic moment of \SI{1}{}$\mu_B$ for values of $U$ ranging from \SI{0}{} to \SI{3}{\electronvolt}, the electronic structure changes dramatically, in agreement with Ref.~\citenum{Bhattacharjee10}. On 2H-NbSe$_2$, various magnetic configurations could be remarkably stabilized with a non-trivial impact of electronic correlations. The latter dramatically change the nature of the magnetic ground states. For instance, without $U$, three types of states are found: low, regular and high spin states (respectively denoted as LS, RS, and HS), with the low spin state being the ground state. A finite $U$ of \SI{2}{\electronvolt} disfavors ``weak'' magnetism and drives the HS state to the ground state. The electronic structure and theoretical STM spectra are found distinct, which could provide a way to better characterize experimentally the magnetic states of the molecules and help guide future ab-initio simulations in choosing the right ingredients to describe the CoPc molecule on 2H-NbSe$_2$.

\section{Method and computational details}
\begin{table*}[ht!]
	\centering
	\setlength\tabcolsep{4.5pt}
	\renewcommand{\arraystretch}{1.5}
	\begin{tabular}{ c | c c c  c}
		 \multicolumn{1}{c|}{{Value of $U$ }} & \multicolumn{4}{c}{{Spin magnetization ($\mu_B$)}}  \\
		 {{(eV)}} & Co & Ligands total & Ligands FM & Ligands AFM  \\
		\hline
	 U=0 & 1.01&-0.11&0.00&-0.12\\
	 U=1 & 1.01&-0.08&0.05&-0.13\\
     U=2 & 1.08&-0.06&0.03&-0.08\\
	 U=3 & 1.08&-0.08&0.03&-0.12
	\end{tabular}
	\caption{Spin magnetization of the free standing CoPc molecule decomposed into the contributions of Co atom and of the ligands. The latter can carry a spin moment that is antiferromagnetically (AFM) or ferromagnetically (FM) coupled to the Co moment. Most of the ligands couple AFM to Co. }\label{Tab:free}
\end{table*}

The simulations were based on density functional theory (DFT) within the plane-wave implementation of the Quantum Espresso package~\cite{QE2009,QE2017}. We used pseudopotentials from the PSLibrary~\cite{DalCorso2014} and the projector augmented wave (PAW) method~\cite{PhysRevB.50.17953}, with the plane waves energy cutoff set to \SI{400}{\rydberg}. The exchange-correlation effects were treated utilizing the PBE generalized gradient functional~\cite{Perdew:2008}. For the free CoPc molecule, a unit cell of \SI{36}{\angstrom}$\times$\SI{36}{\angstrom} was set, and the Brillouin zone sampled with a single k-point. For the geometrical relaxations, a Gaussian smearing of \SI{0.02}{\rydberg} and a convergence threshold of \SI{1e-3}{\atomicunits} was established. For the molecule on the surface, an \SI{8}{}$\times$\SI{8}{} ($\sim$\SI{28}{\angstrom}$\times$\SI{28}{\angstrom}) of the unit cell of 2H-NbSe$_2$ monolayer was constructed with van der Waals  interactions taken into account in the atomic relaxation procedure~\cite{Grimme10}. We follow the experimental observation that Co atom is atop Se~\cite{Wang20}. Electronic correlation effects were accounted for utilizing $U$ values of \SI{1}{}, \SI{2}{} and \SI{3}{\electronvolt}~\cite{Cococcioni05} on the Co atom. The theoretical scanning tunneling microscopy spectra were obtained on the basis of the Tersoff-Hamann model~\cite{Tersoff1983}.

\section{Results and discussions}

\subsection{Free standing CoPc molecule}
\begin{figure*}[ht!]
	\includegraphics[width=0.8\linewidth]{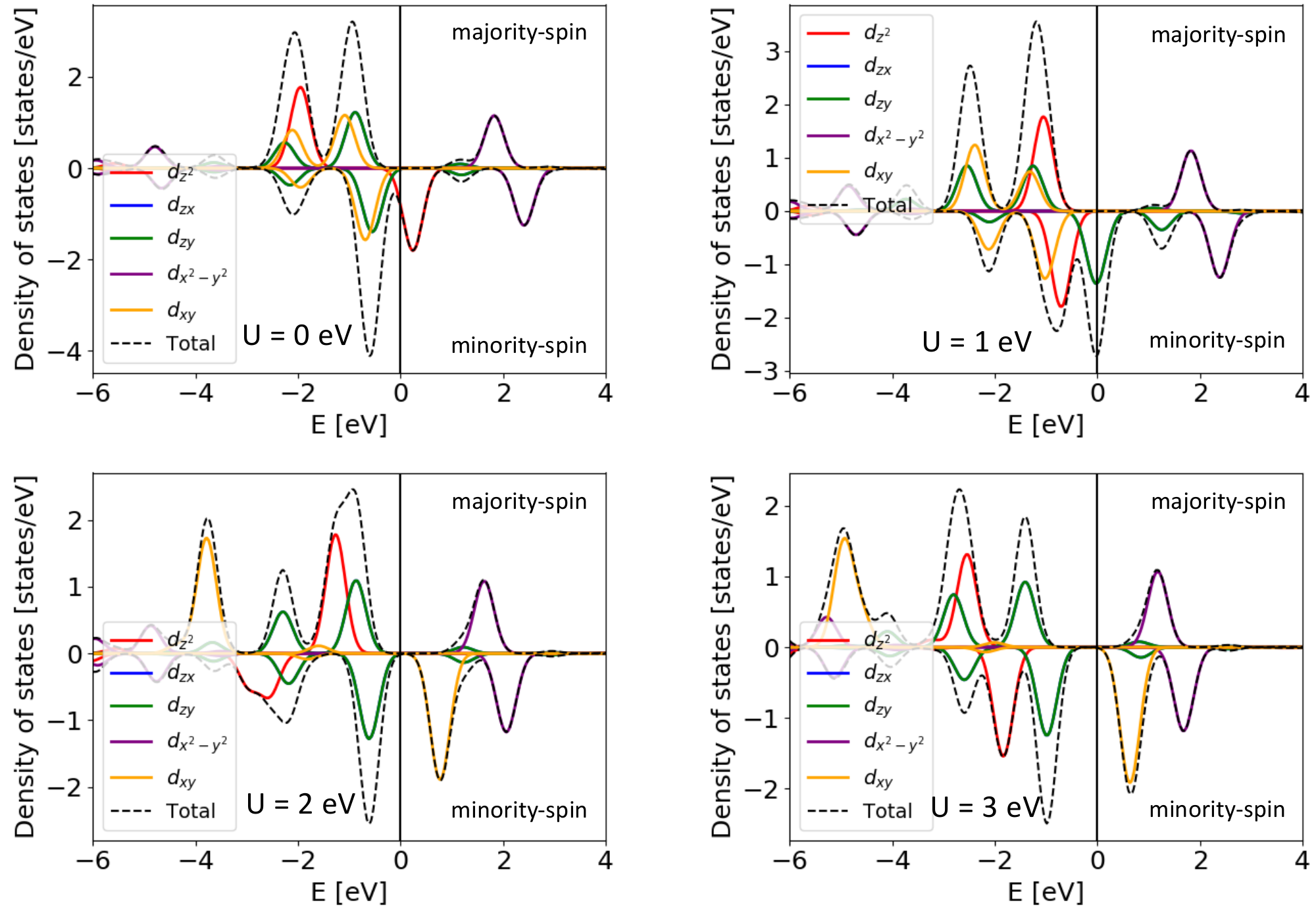}
	\caption{Local density of states (d-states) of the Co atom in the free standing CoPc molecule as function of $U$. The energy scale on the x-axis is defined with respect to the Fermi energy.}\label{Fig:LDOS_free}
\end{figure*}
\begin{figure*}[ht!]
\centering
			\includegraphics[width=7cm]{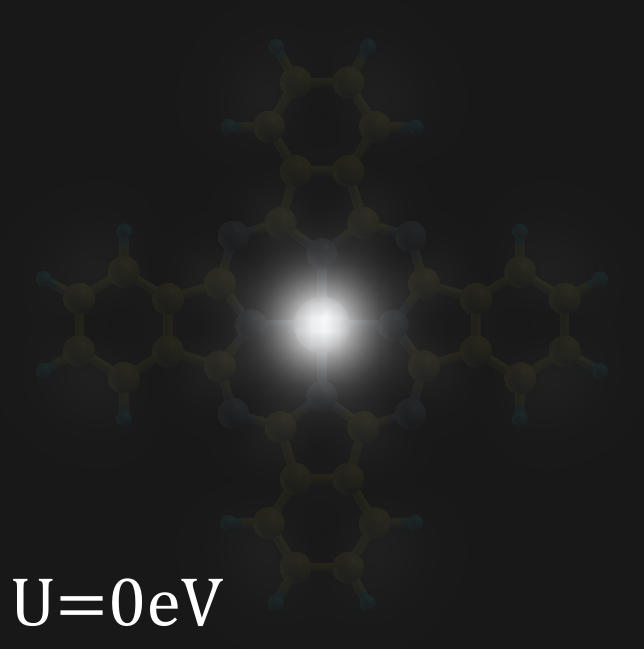}
		\includegraphics[width=7cm]{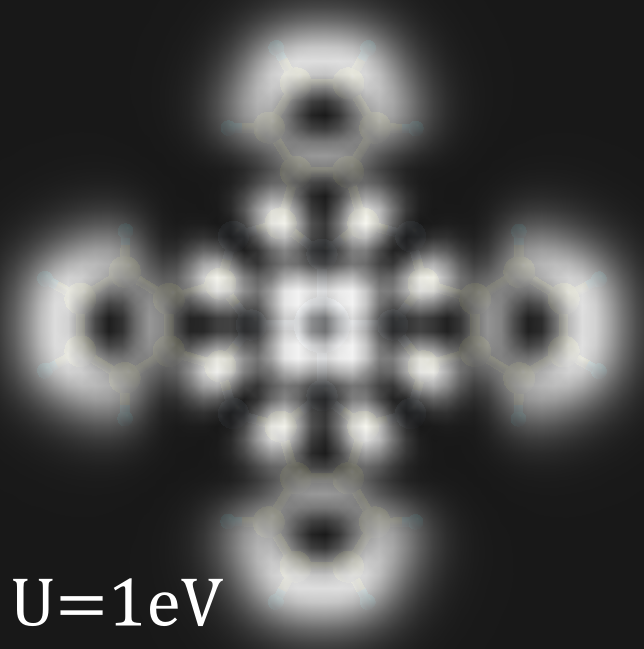}
		\includegraphics[width=7cm]{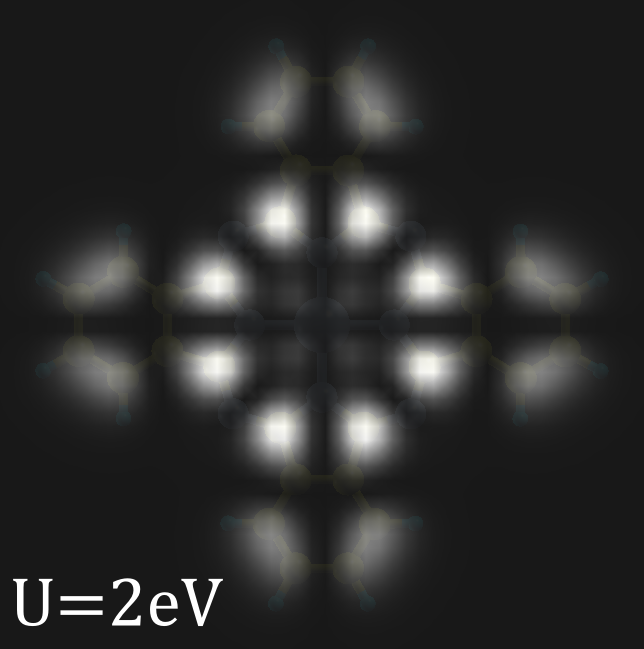}
		\includegraphics[width=7cm]{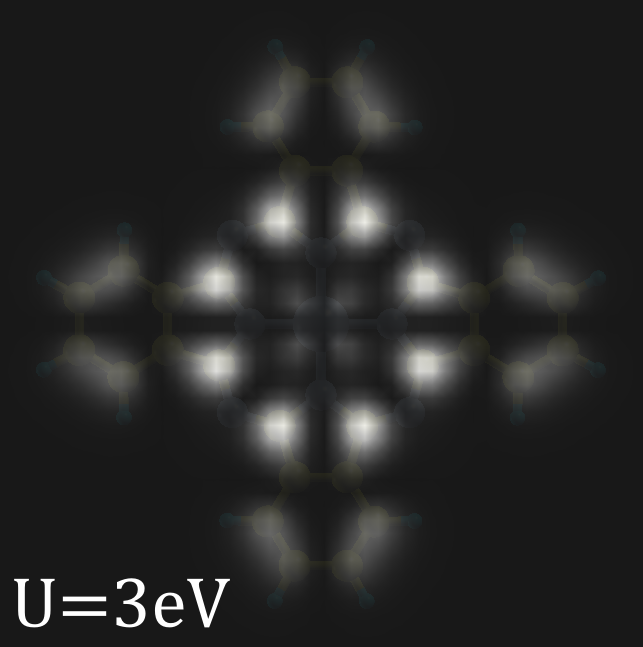}
		\caption{Theoretical STM images obtained at the Fermi energy for the free CoPc molecule for different $U$ values.}\label{Fig:STM_free}
\end{figure*}

When the CoPc molecule is in the gas phase, it carries a magnetic moment of \S{1}{}$\mu_B$ independently of the considered $U$. The spin is thus $1/2$, as known from previous works, while Co is found in the state of Co$^{+2}$ coupled antiferromagnetically to the spin moments of the ligands (see Table~\ref{Tab:free}). The local density of states (LDOS) of the Co atom  for $U = 0$ is shown in Fig.~\ref{Fig:LDOS_free}. One notices that the electronic structure remarkably changes if $U$ is finite, as depicted in Figs.~\ref{Fig:LDOS_free}(b-d). For instance, the partially occupied minority-spin $d_{z^2}$ orbital shifts to lower energies once $U$ is incorporated. In contrast, the  minority-spin $d_{xy}$ orbital  experiences a shift to higher energies upon application of the $U$, and becomes unoccupied. The orbitals that seem to be qualitatively less affected are the $d_{x^2-y^2}$, which remain unoccupied, while the majority-spin $d_{xy}$ orbitals stay occupied. Interestingly the two orbitals $d_{zx}$ and $d_{zy}$, which are degenerate, move to higher energies for for $U = \SI{1}{eV}$, where the corresponding  minority-spinstates reach the Fermi energy, before moving back to initial energy location for larger $U$ values. The characteristics measured in the experimental setups can be compared to the theoretical results to determine the right value of $U$.
The theoretical STM spectra obtained at the Fermi energy are shown in Fig.~\ref{Fig:STM_free}. The electronic features are quite different, as expected, not only at the level of the Co atom (where the $U$ is applied), but also at the ligands. Without $U$, the Co $d_{z^2}$ orbital at the Fermi energy leads to a strong isotropic feature, which dominates over the ligands states. The opposite behavior is found for the other values of $U$ since the $d_{z^2}$ becomes fully occupied leaving the stage for either the $d_{xy}$ orbital for $U = \SI{1}{eV}$ or for a gap when $U = \SI{2}{}$ or $\SI{3}{eV}$. In the latter case, the ligand states are more visible.

\subsection{\texorpdfstring{CoPc/NbSe$_2$}{TEXT}}

\begin{figure*}[ht!]
\centering
	\includegraphics[width=7cm]{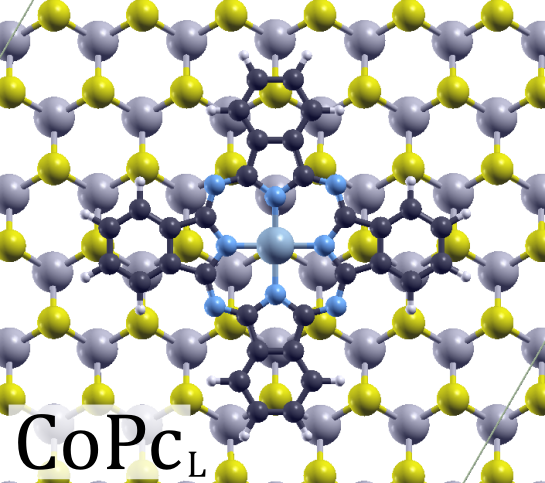}
	\includegraphics[width=7cm]{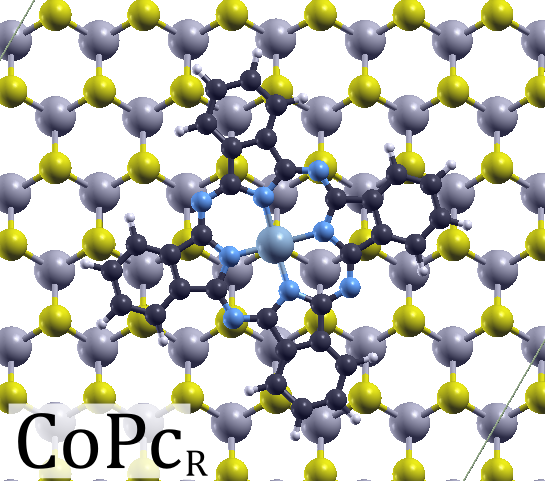}
\caption{CoPc molecule deposited on 2H-NbSe$_2$, where the Co atom is sitting on top of the Se atoms. As indicated by the experiment reported in Ref.~\cite{Wang20}, the molecule can be found in two configurations left (L)  and right (R), which are rotated by 15 degrees from each other, thereby breaking the symmetry of the system in distinct fashion. Two  carbon rings of CoPc$_L$ surround Se atoms in contrast to CoPc$_R$.}\label{Fig:geometry}
	\end{figure*}

The CoPc molecule is deposited on 2H-NbSe$_2$ with the Co atoms sitting atop a Se atom, as found experimentally~\cite{Kezilebieke18,Kezilebieke19,Wang20}. The two configurations identified in Ref.~\citenum{Wang20} are depicted in Fig.~\ref{Fig:geometry}, where the molecules are rotated by \SI{15}{\degree} with respect to each other. The molecule depicted on the left of Fig.~\ref{Fig:geometry} (CoPc$_L$), has two Carbon rings surrounding Se atoms, in contrast to the one on the right (CoPc$_R$). The two molecules break the symmetries of the system in different ways, which can lead to distinct electronic and magnetic behaviors, as discussed below for $U =\SI{0}{}$ and \SI{2}{\electronvolt}.

\subsubsection{\texorpdfstring{Magnetic states of CoPc$_L$ and CoPc$_R$ with $U = \SI{0}{eV}$ }{TEXT}}

Considering various initial input magnetic states, we found three converged magnetic solutions for CoPc$_R$  with $U = \SI{0}{}$. The molecules are found with low, regular or high magnetic state (see Table~\ref{Tab:U=0}). For the three cases, the Co atom is in a 3$^+$ state and couples ferromagnetically to the total spin moment of the ligands, in contrast to what has been observed for the free standing molecule. In the LS state, the Co atom carries a spin moment of \SI{0.11}{}$\mu_B$ while the ligands carry a rather weak spin moment of \SI{0.03}{}$\mu_B$ giving a total of \SI{0.14}{}$\mu_B$. In the RS  state, the Co atom recovers a spin moment of \SI{0.72}{} $\mu_B$ while the ligands have a relatively large moment of \SI{0.19}{}$\mu_B$, adding up to \SI{0.92}{}$\mu_B$, rather close to the moment known for the free standing molecule. For the HS state, Co has a substantial  moment of \SI{1.25}{}$\mu_B$ with a reduced moment of the ligands in comparison to the regular spin state, \SI{0.12}{}$\mu_B$. The total spin moment is thus rather large, \SI{1.37}{}$\mu_B$. 
The magnitude of the spin moment obtained is the three spin states match those detected experimentally (S = 0, 1/2 and 1) considering the g-factor of 1.5. When comparing the energies of the different states, listed in Table~\ref{Tab:U=0}, we surprisingly find that the low-spin state is the ground state with a very large energy difference with respect to the other magnetic states. 
\begin{table*}[ht!]
	\centering
	\setlength\tabcolsep{4.5pt}
	\renewcommand{\arraystretch}{1.5}
	\begin{tabular}{c| c |  c c c  c|c}
 Type of 	& Type of 	 &  \multicolumn{4}{c|}{{Spin magnetization} ($\mu_B$)} & Energy difference \\
molecule	&{spin state}	&  Co & Ligands total &  Ligands FM & Ligands AFM & $\Delta E$ (eV)\\
		\hline
	CoPc$_R$	&LS & 0.11& 0.03 & 0.04&-0.01& 0\\
	CoPc$_R$	&RS & 0.72 & 0.20 & 0.20 & -0.01 & 0.677\\
	CoPc$_R$	&HS & 1.25 &0.08& 0.12&-0.03 & 0.898\\
	\hline
   CoPc$_L$ &LS  & 0.29& -0.10 & 0.15&-0.05& 0 \\
   CoPc$_L$ 		&RS & 0.60 & -0.02 & 0.05&-0.07&  0.788 \\
   CoPc$_L$ &HS  & 1.12 & 0.06 & 0.08 & -0.02&  0.538 	
	\end{tabular}
\caption{Magnetic moments of Co and the ligands of CoPc$_R$ and CoPc$_L$ on H2-NbSe$_2$ as obtained for $U=\SI{0}{}$. Three magnetic states are obtained: regular spin state (RS), low spin state (LS) and high spin state (HS). $\Delta E$ corresponds to the energy difference per atoms of the molecule with respect to the ground state, which is the low spin state.
}\label{Tab:U=0}
\end{table*}

Similarly to CoPc$_R$, CoPc$_L$ with $U=\SI{0}{}$ is also found in three magnetic states with the LS state being the ground state. The difference between the two structural configurations lies in the total spin moment of the ligands, which now in the CoPc$_R$ case can either couple antiferromagnetically or ferromagnetically to that of Co. The Co (ligands) moment equals to \SI{0.29}{}$\mu_B$ (\SI{-0.10}{}$\mu_B$), \SI{0.60}{}$\mu_B$ (\SI{-0.02}{}$\mu_B$) and \SI{1.12}{}$\mu_B$ (\SI{0.06}{}$\mu_B$) respectively for the low, regular and high spin states (see Table~\ref{Tab:U=0}).

When comparing the energies of the two LS ground states obtained for both molecular configurations, we find that of CoPc$_R$ is lower in energy than that of CoPc$_L$ by \SI{89}{\milli\electronvolt} per atom of the molecule. These simulations indicate that it is more likely to observe CoPc$_R$ than CoPc$_L$, which was noticed experimentally~\cite{Wang20}. This explains the possibility of switching one configuration to the other using scanning tunneling microscopy~\cite{Wang20}. Moreover, it is also more likely to detect the RS state in CoPc$_R$ than in CoPc$_L$, while the HS state is more likely to occur in the latter one. On the basis of the present simulations, we conjecture that if one relies experimentally on the observation of the YSR state to characterize the magnetic state of these molecules, it is very probable that the RS state for CoPc$_R$ is the one observed by Wang et al.~\cite{Kezilebieke18,Wang20} since it is lower in energy than the HS states. The low spin state potentially does not introduce a YSR state in the superconducting gap and would correspond then to the S=0 molecule identified by Wang et al.~\cite{Wang20}. In contrast to our results, however, CoPc$_L$ is found in the same experiment to trigger a Kondo resonance that splits under the application of a magnetic field --- an indication that it is magnetic. In the next section we explore the impact of electronic correlations on the magnetic behavior of both molecules.

\begin{table*}[ht!]
	\centering
	\setlength\tabcolsep{4.5pt}
	\renewcommand{\arraystretch}{1.2}
		\begin{tabular}{c| c |  c c c  c|c}
 Type of 	& Type of 	 &  \multicolumn{4}{c|}{{Spin magnetization} ($\mu_B$)} & Energy difference \\
molecule	&{spin state}	&  Co & Ligands total &  Ligands FM & Ligands AFM & $\Delta E$ (eV)\\
		\hline
CoPc$_R$ &	RS & 0.73&-0.06&0.01&-0.07&  0.018 \\
CoPc$_R$ &	HS  & 1.40&0.07&0.13&-0.07& 0\\
\hline
CoPc$_L$	& RS & 0.84&-0.01&0.06&-0.08& 0.915 \\
CoPc$_L$	& HS & 1.33 &0.13&0.13&-0.01& 0 
	\end{tabular}
\caption{Magnetic moments of Co and the ligands of CoPc$_R$ and CoPc$_L$ on H2-NbSe$_2$ as obtained with $U = \SI{2}{\electronvolt}$. Two magnetic states are obtained: regular spin state (RS), and high spin state (HS). $\Delta E$ corresponds to the energy difference per atoms of the molecule with respect to the ground state, which is the high spin state.
}\label{Tab:U=2}
\end{table*}
\begin{figure*}[ht!]
\includegraphics[width=0.8\linewidth]{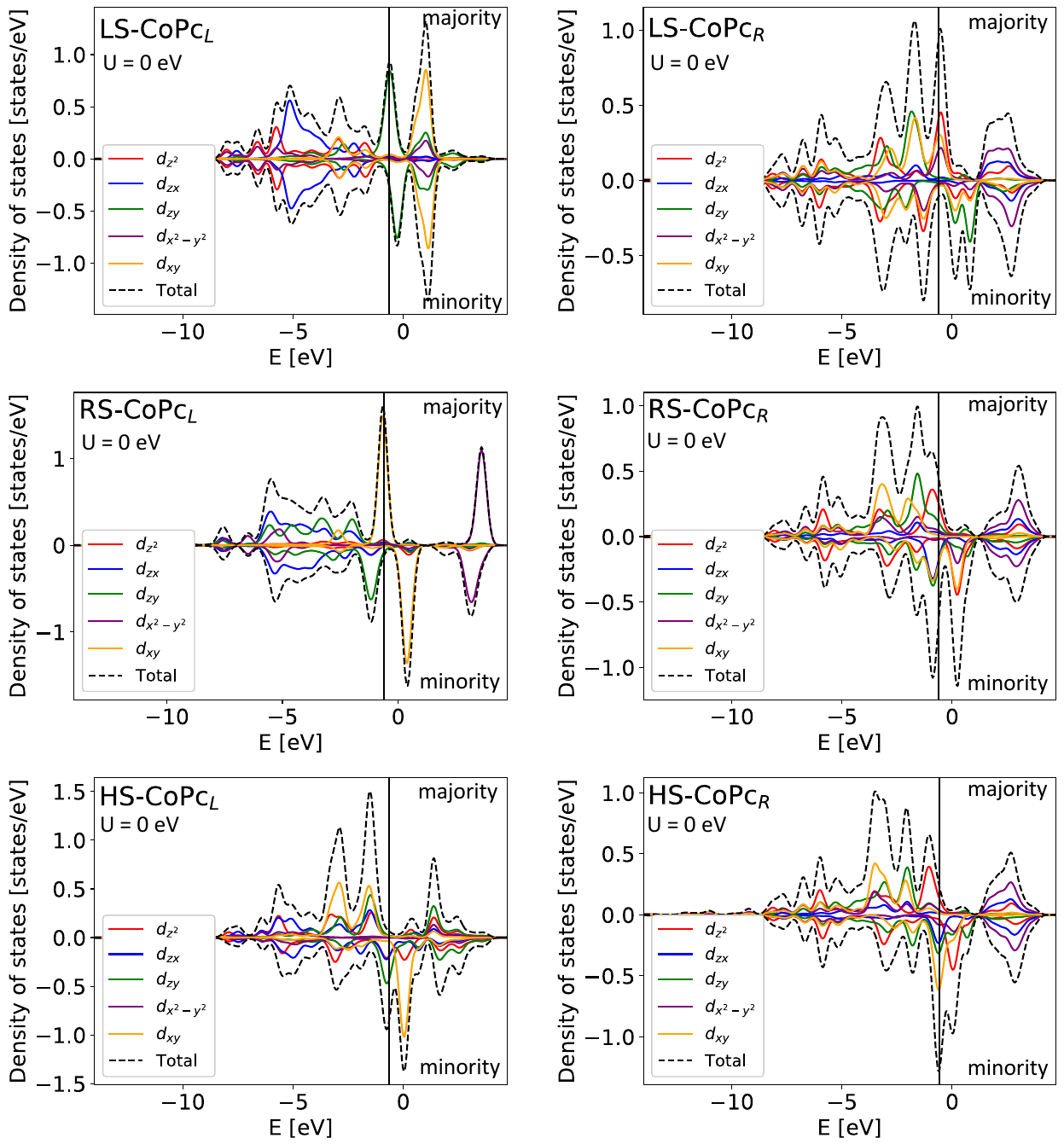}		\caption{Co $d$-resolved local density of states with $U=\SI{0}{\electronvolt}$ of CoPc$_R$  and CoPc$_L$in the three magnetic states: low, regular and high spin states. The energy scale on the x-axis is defined with respect to the Fermi energy.}\label{Fig:LDOS_U=0}
	\end{figure*}
	
\subsubsection{\texorpdfstring{Magnetic states of CoPc$_L$ and CoPc$_R$ with $U = \SI{2}{\electronvolt}$}{TEXT}}
\begin{figure*}[ht!]
\includegraphics[width=0.8\linewidth]{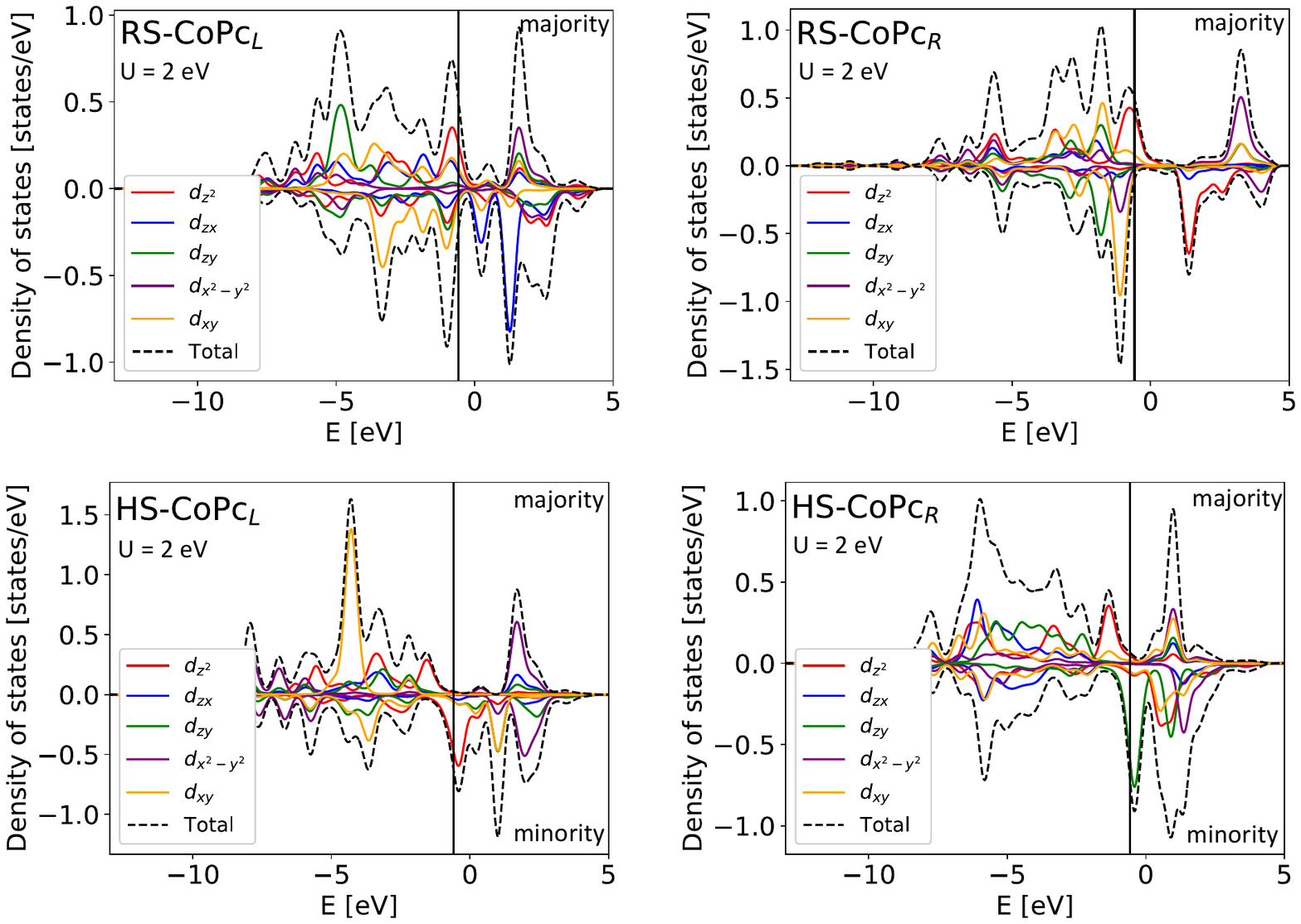}
\caption{Co $d$-resolved local density of states with $U=\SI{2}{\electronvolt}$ of CoPc$_R$ and CoPc$_L$ in the two magnetic states: regular and high spin states. The energy scale on the x-axis is defined with respect to the Fermi energy.}\label{Fig:LDOS_U=2}
	\end{figure*}
	
To investigate the potential impact of correlations, we choose a value of $U = \SI{2}{\electronvolt}$ applied on the Co atom. As already observed in the free standing case,  the electronic correlations dramatically impact the magnetic behavior of the molecules. For instance, the two structurally different molecules are found in RS and HS states but we note that we could not converge LS states (see Table~\ref{Tab:U=2}). Thus, $U$ seems to unfavor the existence of a LS state molecule and promotes the HS state as a ground state over the regular spin state. 

In the RS state, the Co (ligands) spin moment equals \SI{0.73}{}$\mu_B$ (\SI{-0.06}{}$\mu_B$) and \SI{0.84}{}$\mu_B$ (\SI{-0.01}{}$\mu_B$) respectively for CoPc$_R$ and CoPc$_L$. In the HS state, the ligands couple overall ferromagnetically to the Co moment. It is characterized by a Co (ligands) spin moment of \SI{1.40}{}$\mu_B$ (\SI{0.07}{}$\mu_B$) and \SI{1.33}{}$\mu_B$ (\SI{0.13}{}$\mu_B$) for CoPc$_R$ and CoPc$_L$, respectively. Energetically, the two molecules are rather different, even though both of them prefer to be in a HS state. In contrast to the case $U = 0$, the simulations indicate that with $U = \SI{2}{\electronvolt}$, it is more probable to find the molecule CoPc$_L$, which is favored by \SI{0.856}{\electronvolt} per atom of the molecule. For CoPc$_R$, the energy of the HS state is much closer to that of the RS state (\SI{18}{\milli\electronvolt}) in contrast to the \SI{915}{\milli\electronvolt} separating the two magnetic states of CoPc$_L$. In other words, we would expect a higher probability to experimentally observe the RS and HS states for CoPc$_R$, while CoPc$_L$ should be in the HS state. 

\begin{figure*}[ht!]
\centering
\includegraphics[width=7cm]{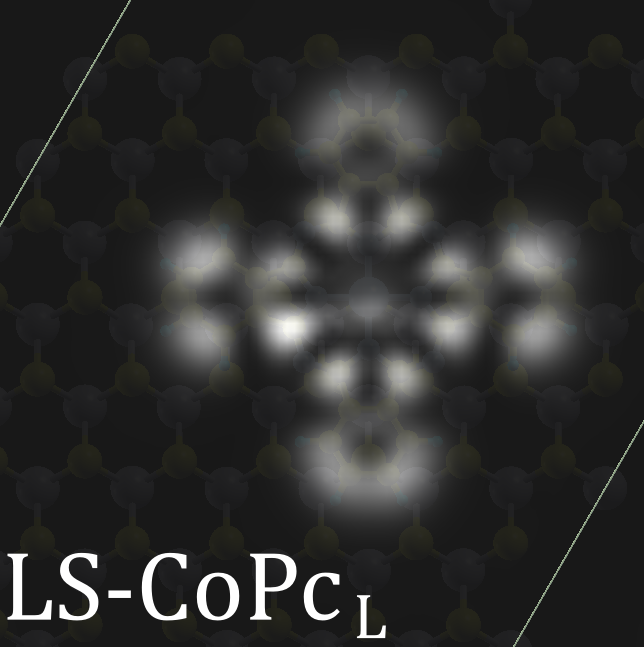}
\includegraphics[width=7cm]{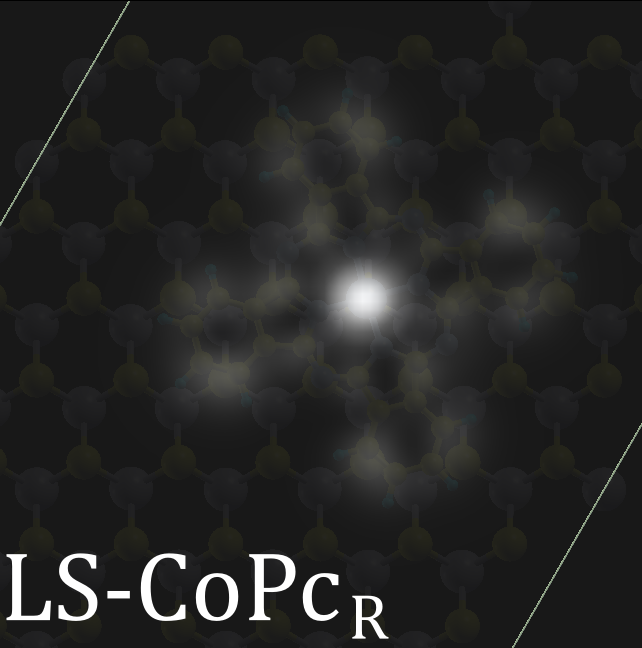}
\includegraphics[width=7cm]{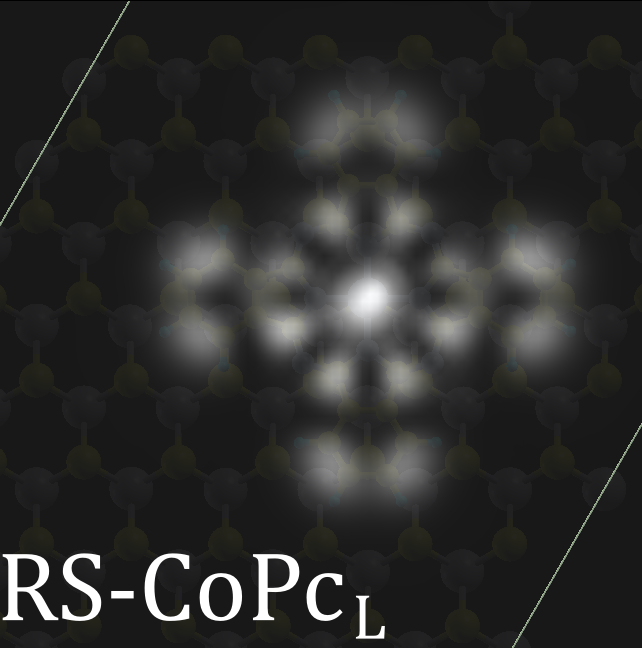}
\includegraphics[width=7cm]{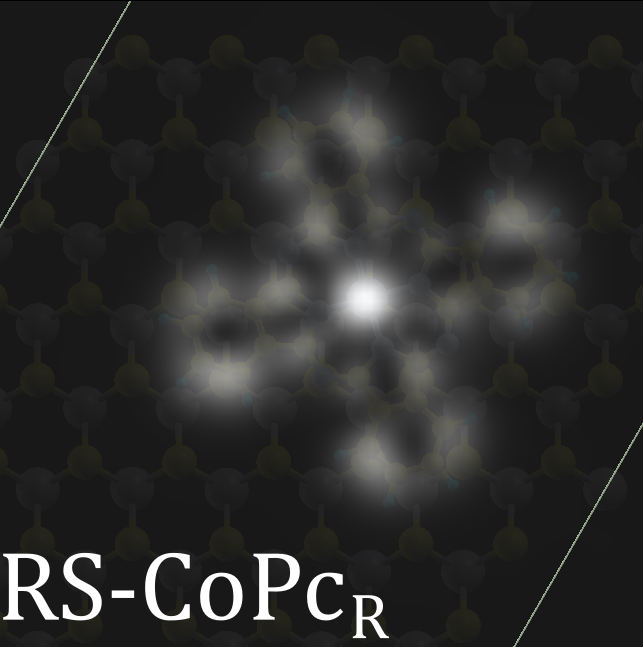}
\includegraphics[width=7cm]{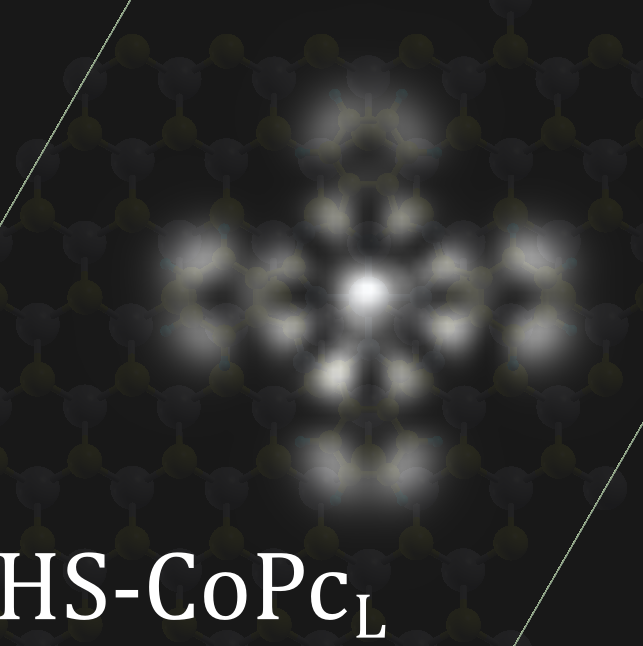}
\includegraphics[width=7cm]{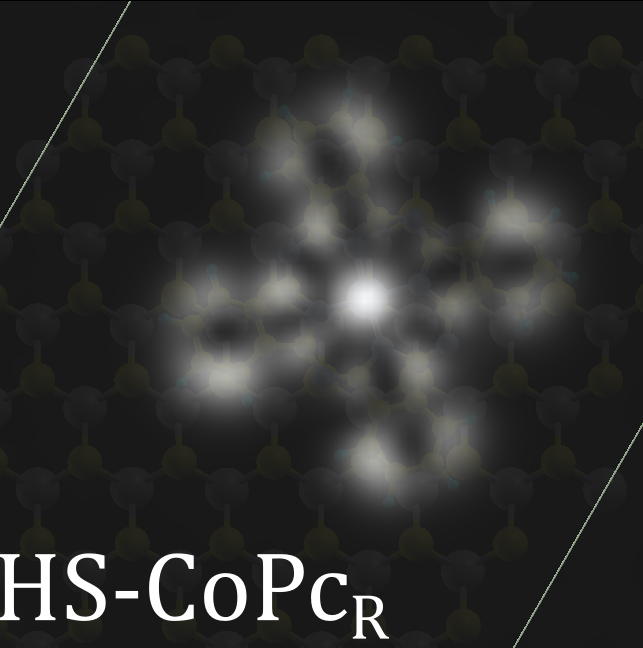}
		\caption{Theoretical STM images of  CoPc$_R$  and CoPc$_L$ molecules on 2H-NbSe$_2$ calculated at the Fermi energy and obtained with $U = \SI{0}{eV}$ for the low, regular and high spin states. }\label{Fig:STM_U=0}
	\end{figure*}

The picture is thus very different from what we obtain without $U$.  Experimentally, the two types of molecules that were characterized via the YSR states are either magnetic~\cite{Kezilebieke18}  or non-magnetic~\cite{Wang20}. The theoretical results obtained without $U$ are therefore more favorable to address the experimental observations at least for CoPc$_R$, while the results obtained with $U = \SI{2}{\electronvolt}$ are more appropriate for CoPc$_L$. This raises the question, whether in practice the degree of electronic correlations changes by rotating the molecule on the surface. 

\begin{figure*}[ht!]
\centering
\includegraphics[width=7cm]{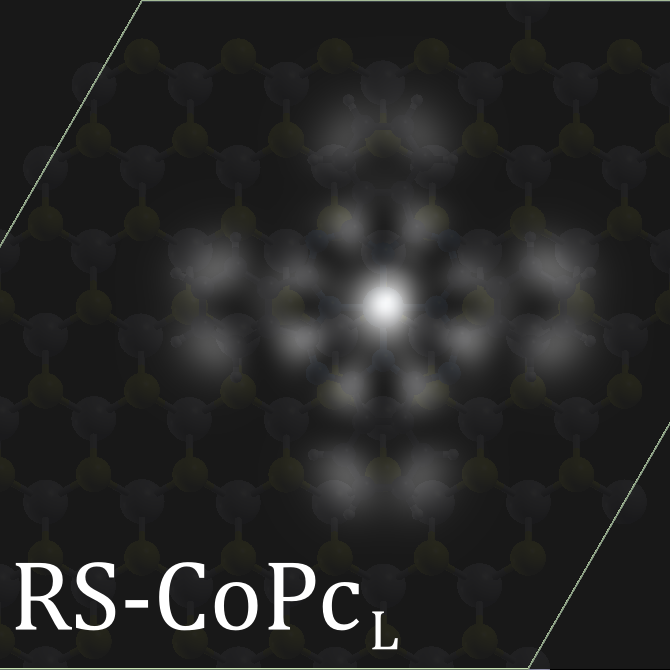}
\includegraphics[width=7cm]{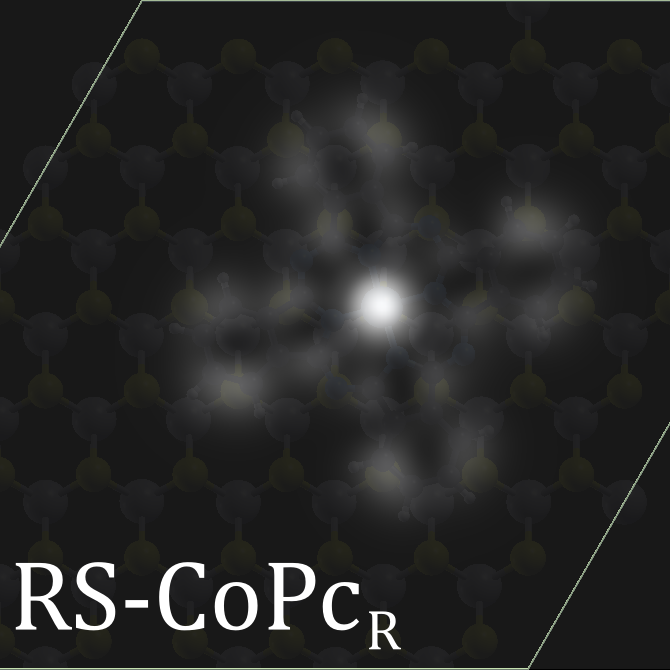}
\includegraphics[width=7cm]{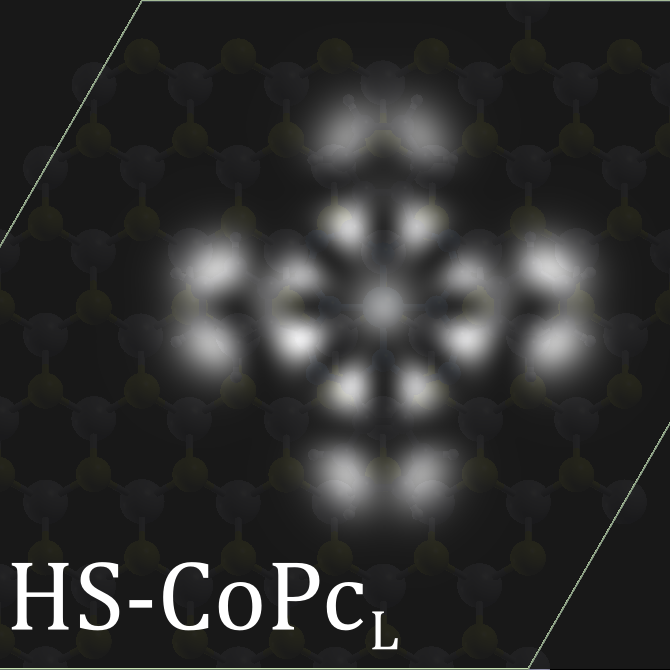}
\includegraphics[width=7cm]{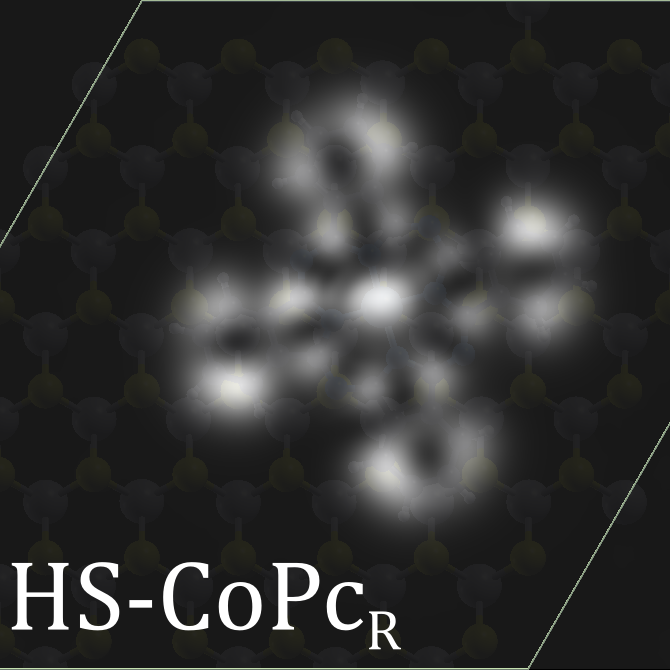}
\caption{Theoretical STM images of  CoPc$_R$  and CoPc$_L$ molecules on 2H-NbSe$_2$ calculated at the Fermi energy and obtained with $U = \SI{2}{eV}$ for the regular and high spin states. }\label{Fig:STM_U=2}
\end{figure*}

\subsubsection{Local density of states and theoretical STM spectra}

The electronic structure encoded in the local density of states of the molecules combined with their theoretical STM/STS spectra could be useful to identify and characterize the experimental findings.

The magnetic state of the molecule as well as its rotation atop of the substrate dramatically modify the electronic structure of the molecule-substrate complex. In Fig.~\ref{Fig:LDOS_U=0}, the d-resolved Co  LDOS for $U = 0$ is plotted for the various  spin states. For CoPc$_L$, the nature of the orbitals lying at the Fermi level changes from the $d_{zy}$ (LS)  to the majority-spin  $d_{xy}$ orbital (RS) and finally a mixture of the minority-spin $d_{xy}$ and $d_{x^2-y^2}$ orbitals (HS). The Co-LDOS is very different for CoPc$_R$. At the Fermi energy of the LS state, the majority-spin $d_{z^2}$ and $d_{x^2-y^2}$ dominate the scene. In addition to the latter orbitals, the minority-spin $d_{zy}$, $d_{zx}$ and $d_{y}$ orbitals get a non-negligible amplitude at the Fermi energy of the RS state.  In contrast to the LS and RS states, the orbitals located at the Fermi energy are mainly of minority-spin nature.

The d-resolved LDOS of the Co atom when $U=\SI{2}{\electronvolt}$ is shown in Fig.~\ref{Fig:LDOS_U=2}.
Similarly to the case with $U=0$, the electronic states change considerably once the molecule is deposited on the surface of 2H-NbSe$_2$.  
The LDOS also depends dramatically on the magnetic state of the molecule, as well as its positioning on the substrate. Starting with the RS state, we find that CoPc$_L$ has a highly intense unoccupied $d_{zx}$ state, which indicates that they are less prone to hybridization than in CoPc$_R$. In contrast, the minority-spin unoccupied $d_{xy}$ state has a rather weak amplitude. At the Fermi energy, both molecules have a rather prominent $d_{z^2}$ state, which has the right symmetry to decay further into the vacuum and therefore to be detected via STM/STS. In the HS state, CoPc$_R$ shows a large $d_{zy}$ state at the Fermi energy in contrast to the $d_{z^2}$ state characterizing CoPc$_L$.

After having discussed the Co LDOS, it is interesting to visualise the theoretical STM spectra calculated at the Fermi energy for the various molecule types with and without $U$. For all cases, the spectra are found very different from those of the free standing molecule. 
Without $U$, the molecule showing the largest (weakest) intensity for Co (ligands) is CoPc$_R$ in the LS state. This is enabled by the large majority-spin $d_{z^2}$ orbital carried by the Co atom (Fig.~\ref{Fig:LDOS_U=0}), which has the right symmetry to be efficiently detected by STM. The absence of that $d_{z^2}$ orbital at the Fermi energy of the molecule LS-CoPc$_L$ makes the feature characterizing Co less intense than the signal emanating from the ligands. For $U = \SI{2}{\electronvolt}$, the $d_{z^2}$ state at the Co atom is isotropic and very intense as illustrated in Fig.~\ref{Fig:STM_U=2} for the three molecules CoPc$_L$ and CoPc$_R$ in their RS states as well as CoPc$_L$ in the HS state. The shape of the feature located at Co in CoPc$_R$ within the HS state is however rather distorted since it mainly originates 
from the $d_{zy}$ state. The intensity of the states emanating from the ligands are less intense than those of Co when the molecules are in the RS state, which provides means to identify whether the molecules are in a RS or a HS state. We note that rotating the molecule on the surface does affect the shape of the ligands states as well as their symmetry, which provides a way of distinguishing between the two types of configurations when utilizing STM.

\section{Conclusion}
To summarize, we performed ab-initio simulations on a CoPc molecule in the free-standing configurations and deposited on a two-dimensional single layer of 2H-NbSe$_2$. We found in the latter case multiple magnetic states with different magnetic moments and electronic structure. Inspired by the recent experimental investigations by Wang et al.~\cite{Wang20}, which showed a dual magnetic behavior of the molecule, we considered two possible orientations of the molecules on the substrate: CoPc$_L$ and CoPc$_R$ atop Se, rotated by 15 degrees from each other. For both geometrical configurations we identified low, regular and high spin states with the low spin state being the ground state when no electronic correlation was added to Co. These correlations where than accounted in the DFT+$U$ framework~\cite{Cococcioni05,Anisimov97} seem to disfavor the low spin state and change the ground state to the high spin state. The energy differences between the different states depend on the way the molecule is oriented on the substrate. The low spin state of CoPc$_R$ obtained without $U$ and the high spin state of CoPc$_L$ with $U = \SI{2}{\electronvolt}$  match the experimental observations. This could advocate for a change of electronic correlations depending on the orientation of the molecule on the surface. Constrained RPA calculations~\cite{Cococcioni05,Aryasetiawan06} may prove useful in this context to evaluate $U$. 
Our work motivates further theoretical and experimental investigations on the unveiled unusual multi-magnetic behavior of the CoPc molecule, which may provide a rich scenario for molecular spintronics applications.

\section*{Acknowledgements}
We acknowledge fruitful discussions with Y. Wang, M. Ternes, N. Lorent\'e, N. Atodiresei and M. Dvorak. Funding is provided by the Priority Programme SPP 2244 2D Materials - Physics of van der Waals Heterostructures of the Deutsche Forschungsgemeinschaft (DFG) (project LO 1659/7-1) and the European Research
Council (ERC) under the European Union's Horizon 2020 research and
innovation programme (ERC-consolidator Grant No. 681405 DYNASORE). 
We gratefully acknowledge the computing time granted by the JARA-HPC Vergabegremium and VSR commission on the supercomputers JURECA at Forschungszentrum J\"ulich and CLAIX at RWTH-Aachen University.

\bibliography{bibliosave}

\end{document}